\documentstyle[12pt,epsfig]{article}
\newcommand{\simlt}
{\mbox{\raisebox{-0.5ex}{$\textstyle \; \sim$}
\raisebox{ 0.8ex}{$\textstyle  \!\!\!\!\!\!\! <$  }}}
\begin{document}
\title{
Unified description of dark matter at the
center and in the halo of the Galaxy}
\author{Neven Bili\'{c}\thanks{Permanent
 address:
Rudjer Bo\v skovi\'c Institute,
P.O. Box 180, 10002 Zagreb, Croatia;
 \hspace*{5mm} Email: bilic@thphys.irb.hr}\,,
Gary B.\ Tupper, and Raoul D.\ Viollier\thanks{
 Email: viollier@physci.uct.ac.za}
\\
Institute of Theoretical Physics and Astrophysics, \\
 Department of Physics, University of Cape Town,  \\
 Private Bag, Rondebosch 7701, South Africa \\
 }
\maketitle
\begin{abstract}
We consider a self-gravitating ideal
fermion gas at nonzero temperature as a model for the
Galactic halo.
The Galactic halo of mass
$\sim 2 \times 10^{12} M_{\odot}$ enclosed within
a radius of $\sim 200$
kpc is consistent with the existence of a
supermassive compact dark object at
the Galactic center that is in hydrostatic and
quasi-thermal equilibrium with the halo.
The central object has a maximal mass of $\sim 2.3 \times 10^{6}
M_{\odot}$ within a minimal radius of $\sim 18$ mpc for
fermion masses $\sim 15$ keV.
\end{abstract}
%\pacs{PACS Nos.:  98.35.Gi, 98.35.Jk, 95.35.+d}
In the past,
self-gravitating neutrino matter was suggested as a model
for quasars,
with neutrino
masses in the range $0.2 {\rm keV} \simlt m \simlt 0.5 {\rm MeV}$
\cite{1}.
More recently,
supermassive compact
objects consisting of nearly non-in\-ter\-act\-ing degenerate
fermionic matter, with fermion masses
in the range $10 \simlt m/{\rm keV}$ $\simlt 20$,
have been proposed
\cite{2,3,4,5,6}
 as an alternative to the
supermassive black holes that are
believed to reside at the centers of many
galaxies.

So far the masses of $\sim 20$ supermassive
compact dark objects at the galactic
centers have been measured
\cite{kor}.
The most massive compact dark object ever observed is
located at the center of M87 in the Virgo cluster,
and  it has a mass of
$\sim 3 \times 10^9 M_{\odot}$ \cite{7}.
If we identify this object
of maximal mass with a degenerate fermion star
at the Oppenheimer-Volkoff (OV) limit \cite{opp},
i.e., $M_{\rm{OV}} = 0.54 M_{\rm Pl}^3\,
m^{-2} g^{-1/2} \simeq 3 \times 10^9
M_{\odot}$
\cite{4},
where $M_{\rm{Pl}}=\sqrt{\hbar c/G}$,
 this allows us to fix the fermion mass to $m \simeq 15$ keV
for a spin and particle-antiparticle degeneracy factor  $g=2$.
Such a relativistic object
would have a radius $R_{\rm{OV}}= 4.45
R_{\rm{S}} \simeq$ 1.5 light-days,
where $R_{\rm{S}}$ is the Schwarzschild
radius of the mass $M_{\rm{OV}}$.
It would thus be virtually indistinguishable from
a black hole of the same mass,
as the closest stable orbit around a black hole
has a radius of 3 $R_{\rm{S}}$ anyway.

Near the lower end of the observed mass  range
is the compact dark object
located at the Galactic center
\cite{eck}
 with a mass $M_{\rm c} \simeq 2.6 \times
10^6 M_{\odot}$.
Interpreting this object as a degenerate fermion star consisting of
$m \simeq 15$ keV and $g= 2$ fermions,
the radius is $R_{\rm c} \simeq 21$ light-days
$\simeq 7 \times 10^4 R_{\rm S}$
\cite{2},
$R_{\rm S}$ being the Schwarzschild
radius of the mass $M_{\rm c}$.
Such a nonrelativistic object is far from being a black hole.
The observed motion of stars within a projected distance of
$\sim$ 6 to $\sim$ 50 light-days from Sgr A$^{*}$
\cite{eck}
yields, apart from the
mass, an upper limit for
the radius of the fermion star $R_{\rm c}  \simlt 22$
light-days.

The required nearly
non-interacting fermion of $\sim$ 15 keV mass cannot be an
active neutrino,
as it would overclose the Universe by orders of magnitude
\cite{11}.
However, the $\sim\! 15$ keV fermion could very well be a sterile
neutrino mixed to active neutrinos with a mixing angle
$\sin^{2} 2 \theta \sim 10^{-11}$.
%$\Omega_{\rm d} \simeq$ 0.3 to the dark matter
%fraction of the critical density today.
Indeed, as has been shown for an
initial lepton asymmetry of
$\sim 10^{-2}$,  a sterile neutrino of mass
$\sim$ 10 keV may be
resonantly or non-resonantly produced in the early Universe with near
closure density, i.e., $\Omega_{\rm d} \sim$ 0.3
\cite{shi}.
As an alternative possibility,
the required $\sim$ 15 keV
 fermion could be the axino
\cite{cov}
or the gravitino
\cite{lyt}
in soft supersymmetry breaking scenarios.

In the recent past,
galactic halos were successfully modeled as
a self-gravitating
isothermal gas of particles of arbitrary mass,
the density of which  scales asymptotically as
$r^{-2}$, yielding flat rotation curves
\cite{col}.
As the supermassive compact
dark objects at the galactic centers
are well described by a gas of fermions
of mass $m \sim 15$ keV at $T = 0$,
it is tempting to explore the
possibility that one could describe
both the supermassive compact dark objects
and their galactic halos in
a unified way in terms of a fermion gas
at finite temperature.
We  show in this letter that this
is indeed the case, and that
the observed dark matter distribution in the
Galactic halo is consistent with
the existence of a supermassive compact dark
object at the center of
the Galaxy which has about the right mass and size.

Degenerate fermion stars are well understood
in terms of the Thomas-Fermi theory applied
to self-gravitating fermionic matter at $T = 0$
\cite{2}.
Extending this theory to
nonzero temperature
\cite{16,bil1,bil2,17},
it has been shown
that at some critical
temperature $T = T_{\rm c}$,
a self-gravitating ideal fermion gas, having a mass
below the  OV limit  enclosed in a
sphere of radius $R$, may
undergo a first-order gravitational
phase transition from a diffuse state to a
condensed state.
However, this first-order
phase transition can take place only
if the Fermi gas is able to get rid
of the large latent heat.
As short-range interactions
of the fermions are negligible, the gas cannot
release its latent heat;
it will thus be trapped for temperatures $T <
T_{\rm c}$ in a thermodynamic
quasistable supercooled state
close to the point of gravothermal
collapse.

The formation of a supercooled state
close to the point of
gravothermal collapse
may be due to
violent relaxation
\cite{lynd,shu,kul}.
Through the gravitational collapse of an
overdense fluctuation, $\sim$ 1 Gyr after the Big Bang,
part of gravitational energy transforms into the kinetic energy
of random motion of small-scale density fluctuations.
The resulting virialized
cloud will thus be well
approximated by a gravitationally stable quasi-thermalized
halo.
In order to
estimate the particle mass-temperature ratio,
we assume that a cold overdense cloud of
the mass of the Galaxy $M$,
 stops expanding at the time $t_{\rm m}$,
reaching its maximal radius
$R_{\rm m}$ and minimal
average density $\rho_{\rm{m}}= 3 M/(4 \pi R_{\rm{m}}^3)$.
 The total energy per particle is just the gravitational energy
 \begin{equation}
 E=-\frac{3}{5}\frac{GM}{R_{\rm{m}}} \, .
 \label{eq001}
 \end{equation}
 Assuming spherical collapse
 \cite{padma},
 one arrives at
 \begin{equation}
 \rho_{\rm{m}}=\frac{9\pi^2}{16} \bar{\rho}(t_{\rm{m}})
 =\frac{9\pi^2}{16} \Omega_{\rm{d}} \rho_0 (1+z_{\rm{m}})^3,
 \label{eq002}
 \end{equation}
 where $\bar{\rho}(t_{\rm{m}})$ is the background density
 at the time $t_{\rm{m}}$ or the cosmological
 redshift $z_{\rm{m}}$, and
$\rho_0\equiv 3 H_0^2/(8\pi G)$ is the present critical
 density.
 We now
 approximate the virialized cloud by
a singular isothermal sphere
\cite{bin}
of mass $M$ and
radius $R$,
characterized by
a constant circular velocity
$ \Theta=(2 T/m)^{1/2}$
and the density profile
$ \rho(r)=\Theta^2/4\pi G r^2 .$
% \begin{equation}
% \rho(r)=\frac{\Theta^2}{4\pi G r^2}\, .
% \label{eq003}
% \end{equation}
Its total energy per particle is the sum of gravitational
and thermal energies, i.e.,
 \begin{equation}
 E=-\frac{1}{4}\frac{GM}{R}
 =-\frac{1}{4}\Theta^2 .
 \label{eq004}
 \end{equation}
 Combining Eqs. (\ref{eq001}), (\ref{eq002}),
 and (\ref{eq004}),
 we find
 \begin{equation}
 \Theta^2=\frac{6\pi}{5} G
 (6 \Omega_{\rm{d}} \rho_0 M^2)^{1/3}(1+z_{\rm{m}}) .
 \label{eq005}
 \end{equation}
 Taking $\Omega_{\rm{d}}=0.3$, $M=2\times 10^{12} M_{\odot}$,
 $z_{\rm{m}}=4$, and $H_0=65\,{\rm km\, s^{-1} Mpc^{-1}}$, we find
 $\Theta \simeq 220\, {\rm km\, s^{-1}}$, which corresponds to the
 mass-temperature ratio $m/T\simeq 4\times 10^6$.

We now briefly discuss the
Tho\-mas-Fer\-mi theory
\cite{bil1,bil2}
for a self-gravitating gas
of $N$ fermions with mass $m$
at the temperature $T$ enclosed in a sphere of radius $R$.
We restrict our attention to
the  Newtonian theory
since the general relativistic effects
are not relevant to the physics we discuss
in this paper.
The general relativistic treatment will be reported elsewhere.
For large $N$, we can
assume that  fermions move in a spherically
 symmetric
 mean-field potential
$\varphi (r)$ which satisfies Poisson's equation
\begin{equation}
\frac{d\varphi}{dr}=G \frac{{\cal M}}{r^2} \, ;
\;\;\;\;
\frac{d{\cal{M}}}{dr}=4\pi r^2 m n \, ,
\label{eq47}
\end{equation}
$\cal M$ being the enclosed mass.
The number density  of
fermions (including antifermions) $n$ can be expressed
in terms of the Fermi-Dirac distribution
(in units $\hbar=c=k=1$)
\begin{equation}\label{eq02}
n=\frac{\rho}{m}
 = g \int \frac{d^3q}{(2\pi)^3}\,
\left(1+\exp(\frac{q^2}{2mT}+\frac{m}{T}\varphi
 -\frac{\mu}{T}) \right)^{-1} \, .
\end{equation}
Here
$g$ denotes the combined spin-degeneracy factor of
the neutral
fermions and antifermions, i.e., $g$ is 2 or 4 for Majorana
 or Dirac fermions, respectively.
 For each solution
 $\varphi(r)$ of (\ref{eq47}),
the chemical potential $\mu$ is adjusted so that
 the constraint
\begin{equation}\label{eq04}
\int_0^R dr\,4\pi r^2 n(r)=N,
\end{equation}
is satisfied.
Equations (\ref{eq47}) with (\ref{eq02})
should be integrated using
the bo\-un\-da\-ry conditions at the origin, i.e.,
\begin{equation}
\varphi(0)=\varphi_0
\, ; \;\;\;\;\;
{\cal{M}}(0)=0.
\label{eq03}
\end{equation}
It is useful to introduce the degeneracy
parameter
\begin{equation}
\eta=
 \frac{\mu}{T} -
\frac{m}{T}\varphi \; \; .
\label{eq05}
\end{equation}
As $\varphi$ is monotonously increasing with increasing
 $r$, the strongest degeneracy is obtained at the center
 with $\eta_0=(\mu-m\varphi_0)/T$.
The parameter $\eta_0$,
uniquely related to the central
density,
will eventually be fixed
by the constraint (\ref{eq04}) or equivalently
by the condition
${\cal M}(R)=mN$
at the outer boundary.
In this way,
the explicit dependence on
the chemical potential $\mu$
is absorbed in the degeneracy parameter $\eta_0$.
For $r\geq R$, the function $\varphi$ yields
the usual empty-space Newtonian potential
\begin{equation}
\varphi(r)=- \frac{mN}{r} \, .
\label{eq91}
\end{equation}
The set of self-consistency
equations (\ref{eq47})-(\ref{eq04}), with the boundary
conditions (\ref{eq03}),
defines the gravitational Thomas-Fermi
equation.

The numerical procedure is now straightforward.
For a fixed, arbitrarily chosen
ratio $m/T$,
 we first  integrate
Eqs.
 (\ref{eq47})
 numerically
on the interval $[0,R]$ to find
the solutions
 for various central values
$\eta_0$.
This yields
${\cal M}(R)$ as a function of $\eta_0$.
We then select the
value of $\eta_0$
for which
${\cal M}(R)=mN$.

The quantities
$N$, $T$,and $R$  are free parameters in our model
and their
values are dictated
by  physics.
In the following,  $N$ is required to be
of the order $2\times 10^{12} M_{\odot}/m$,
so that for any $m$, the total mass
is close to the estimated mass of the halo
\cite{wilk}.
As we have demonstrated,
the expected temperature of the halo
is given by $m/T=4\times 10^6$.
Our choice
$R=200$ kpc
is based on the estimated size of the Galactic halo.
The only remaining free parameter is the fermion
mass, which we  fix at
$m=15$ keV, and justify its choice {\em a posteriori}.

\begin{figure}[p]
\centering
\epsfig{file=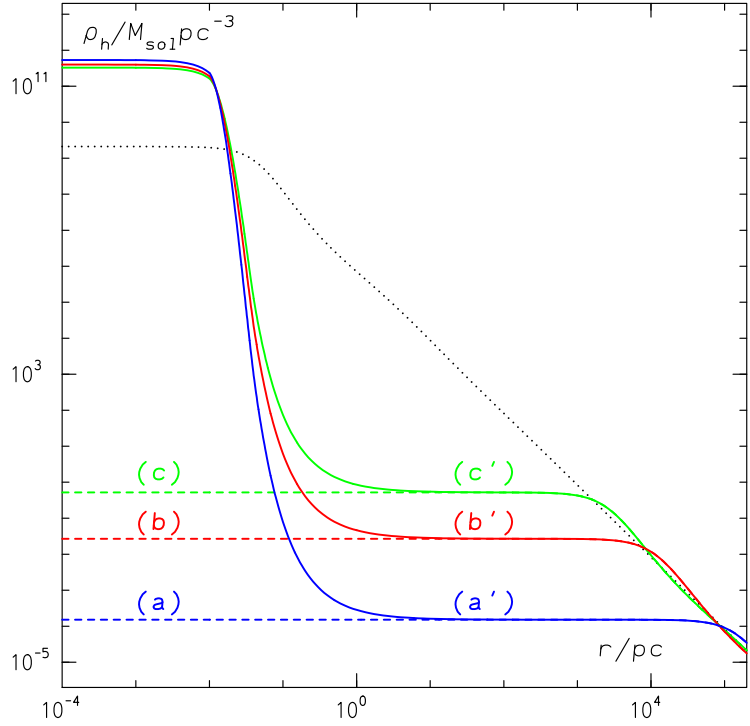,width=10cm}
\caption{
The mass density profile of the halo
for $\eta_0=0$ (dotted line) and for
the six $\eta_0$-values discussed in the text.
Configurations with negative $\eta_0$
((a), (b), (c)) are depicted by the dashed
and those with positive $\eta_0$
((a'), (b'), (c')) by the solid line.
}
\label{fig1}
\end{figure}

For fixed $N$, there is a range of $T$
where the Thomas-Fermi equation has multiple solutions.
For example, for $N=2\times 10^{12}$ and
$m/T=4\times 10^6$, we find six solutions,
which we denote by
(a), (b), (c), (c'), (b'), and (a')
corresponding to the values $\eta_0 =$
 -30.53,
 -25.35,
 -22.39,
  29.28,
  33.38, and
  40.48, respectively.
In Fig.\ \ref{fig1}
we plot the mass density profiles
of the halo.
For the negative central value $\eta_0$,
for which the degeneracy parameter is negative everywhere,
the system behaves basically as a
Maxwell-Boltzmann isothermal sphere.
Positive values of the central degeneracy parameter $\eta_0$
are characterized by a pronounced central core
of mass of about $2.5 \times 10^6 M_{\odot}$
within a radius of about 20 mpc.
The presence of the core is obviously due to
the degeneracy pressure.
%A similar structure was obtained in
%collisionless stellar systems modeled as
%a nonrelativistic Fermi gas
%\cite{chav}.
The core represents material which, having been
cooled by expansion, experiences little entropy
increase during the ensuing collapse. Thus the dynamics of
its formation should be well approximated by
a dynamical Thomas-Fermi theory based on the equation
of state of a degenerate Fermi gas
\cite{bil3}.
%[30]
Conversely, the halo is formed from phase-mixed
matter and estimates similar to those leading to (\ref{eq005}) give
an average entropy
per particle increasing from {\it few} $\times$ 10$^{0}$ to
{\it few} $\times$ 10$^{1}$.

A similar structure was obtained in collisionless stellar systems
modeled as a nonrelativistic Fermi gas \cite{chav}. Note that
while violent relaxation leads to a Fermi-Dirac distribution
in either case, for stars the onset of degeneracy signals the
breakdown of the assumption that collisions are unimportant, resulting
in a Maxwell-Boltzmann distribution
\cite{shu}.
No such breakdown occurs for elementary
fermions
\cite{kul}.

Fig.\ \ref{fig1}
shows two important features.
First,
a galactic halo at a given temperature
may or may not have a central core
depending on whether  the central degeneracy parameter $\eta_0$
is positive or negative.
Second,
the closer to zero $\eta_0$ is,
the smaller the radius at which the
$r^{-2}$  asymptotic behavior of  density begins.
The flattening of the Galactic rotation curve
begins in the range $1 \simlt r/{\rm kpc} \simlt 10$,
hence the solution (c') most likely describes the
Galactic halo.
This may be verified by calculating the rotation
curves in our model.
We know already from our estimate (\ref{eq005})
that our model
yields the correct asymptotic circular velocity of
220 km/s.
In order to make a more realistic comparison
with the observed Galactic rotation curve,
we must include
two additional matter components: the bulge and
the disk.
The bulge is modeled as a spherically symmetric matter distribution
of the form
\cite{you}
\begin{equation}
\rho_{\rm b}(s)=\frac{e^{-hs}}{2s^3}
\int_0^{\infty} du
\frac{e^{-hsu}}{[(u+1)^8-1]^{1/2}} \, ,
\label{eq006}
\end{equation}
where $s=(r/r_0)^{1/4}$, $r_0$ is the effective radius of the bulge
and $h$ is  a parameter.
We adopt $r_0=2.67$ kpc
and $h$ yielding a bulge mass
$M_{\rm b}= 1.5 \times 10^{10} M_{\odot}$
\cite{suc}.
\begin{figure}[p]
\centering
\epsfig{file=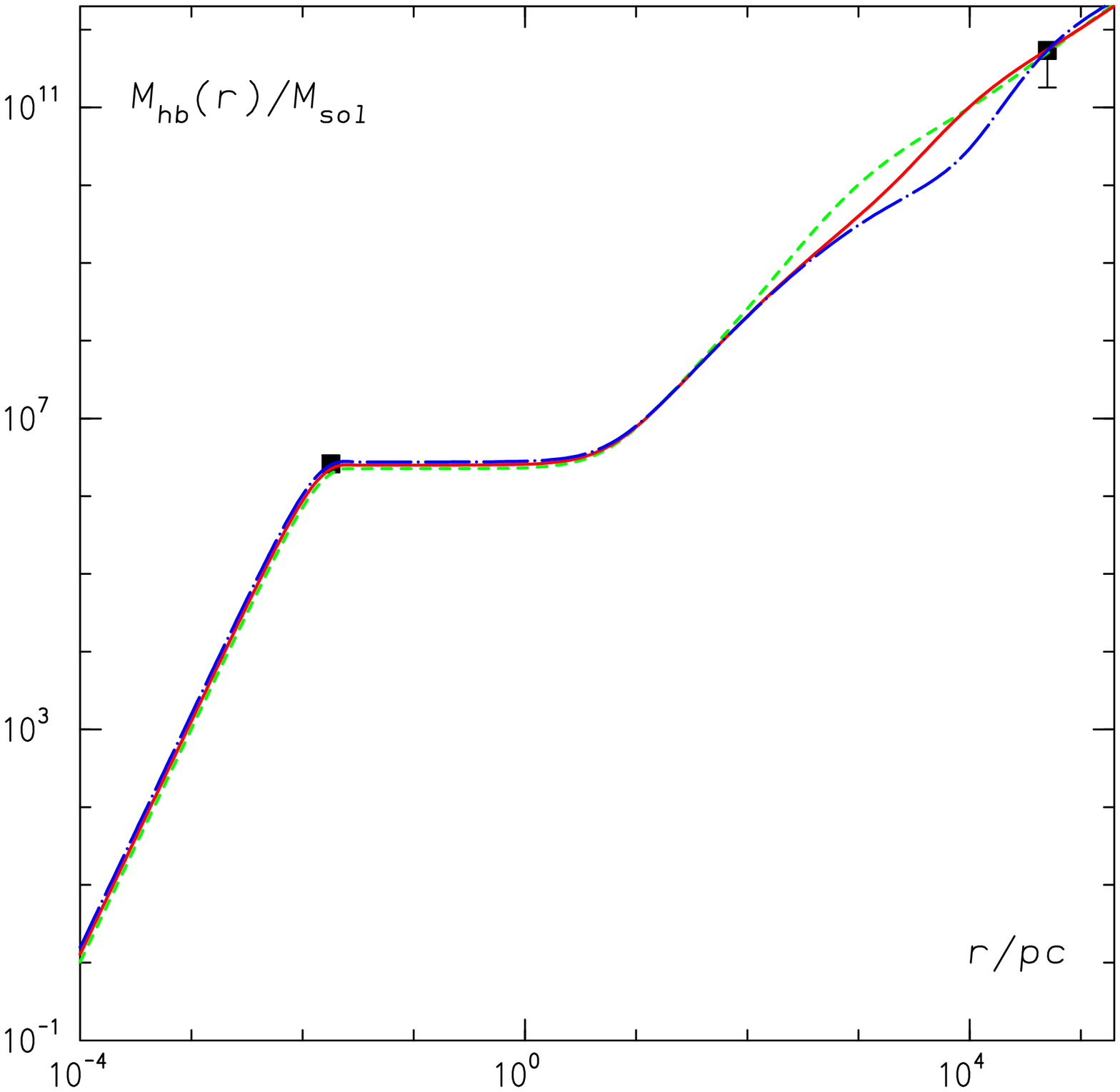,width=10cm}
\caption{
Enclosed mass of the halo plus bulge versus
radius for $\eta_0$ =
   24 (dashed),
  28 (solid),
and
    32  (dot-dashed line).
}
\label{fig2}
\end{figure}
In Fig.\ \ref{fig2} the mass
of the halo and bulge enclosed within
a given radius is plotted for various $\eta_0$.
The data points, indicated  by squares, are
the  mass
$M_{\rm c}=2.6 \times 10^6 M_{\odot}$ within
18 mpc, estimated from the motion of the stars
near Sgr A$^*$ \cite{eck},
and the mass
$M_{50}=5.4^{+0.2}_{-3.6}\times 10^{11}$
within 50 kpc,
 estimated from
the motion
of satellite galaxies and globular clusters
\cite{wilk}.
Variation of the central degeneracy parameter
$\eta_0$ between 24 and 32 does not change
the essential halo features.

\begin{figure}[p]
\centering
\epsfig{file=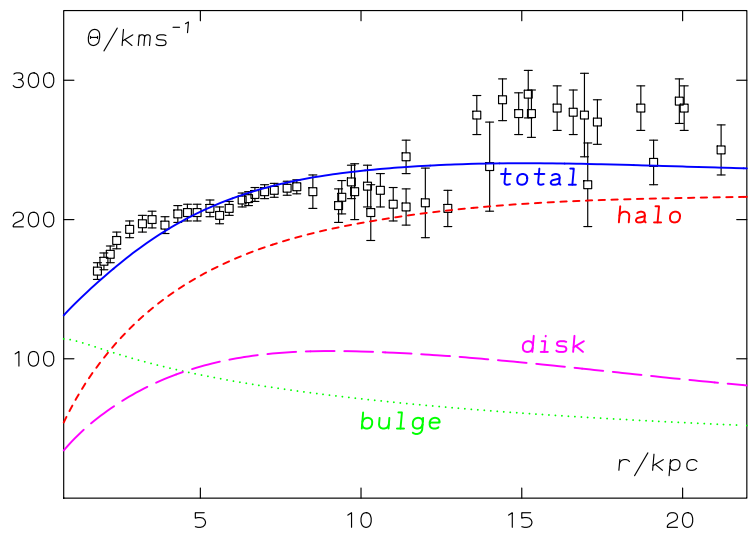,width=10cm}
\caption{
Fit to the Galactic rotation curve.
The data points are
by Olling and Merrifield \protect\cite{oll},
for $R_0=8.5$ kpc and $\Theta_0=220$ km/s.
}
\label{fig3}
\end{figure}

In Fig.\ \ref{fig3} we plot
the circular velocity components:
the halo, the bulge, and the disk.
The contribution of the disk
is modeled  as \cite{per}
\begin{equation}
\Theta_{\rm d}(r)^2=
\Theta_{\rm d}(r_{\rm o})^2
\frac{1.97 (r/r_{\rm o})^{1.22}}{
[(r/r_{\rm o})^2+0.78^2]^{1.43}} \, ,
\label{eq007}
\end{equation}
where we take
$r_{\rm o}=13.5$ kpc and
$\Theta_{\rm d}=100$ km/s.
Here we have assumed for simplicity that
the disc does not influence the mass distribution
of the bulge and the halo.
Choosing the central degeneracy
$\eta_0=28$ for the halo, the data
by Merrifield and Olling \cite{oll} are reasonably well
fitted.

We now turn to the discussion of
our choice of the fermion mass $m=15$ keV
for the degeneracy factor $g=2$.
To that end,
we  investigate how  the mass of the
central object,
i.e., the mass $M_{\rm c}$ within 18 mpc,
depends on $m$ in the interval
5 to 25 keV,
for various
$\eta_0$.
We find that $m\simeq15$ keV always gives the maximal value of
$M_{\rm c}$
ranging between 1.7 and 2.3 $\times 10^6 M_{\odot}$
for $\eta_0$ between 20 and 28.
Hence, with  $m\simeq 15$ keV  we get the
value closest to the mass of the central object
$M_{\rm c}$
estimated from the motion of the stars
lear Sgr A$^*$ \cite{eck}.

%%%%%%%PAGE 9

The radius of our central object of about 18 mpc is much larger than
the size of
the radio source Sgr A$^{*}$. In fact, very large array interferometric observations
of Sgr A$^{*}$ at millimeter wavelength show that the radiowave emitting
region is $\leq$ 1-3 AU \cite{lo}.
However,
it has not yet been shown conclusively
that Sgr A$^{*}$ is indeed the object that has
a mass $\sim$ 3 $\times$ 10$^{6}$ M$_{\odot}$.
There are arguments, based on the nonmotion of Sgr A$^{*}$ and
equipartition of energy in the central star cluster,
indicating that Sgr
A$^{*}$ could have a mass $\geq$ 10$^{3}$ M$_{\odot}$ \cite{zha}.
This argument is only conclusive if equipartition of energy actually
takes place in a reasonable time frame. It is, therefore, still possible that the
compact radiosource Sgr A$^{*}$,
with a radius of a few AU, and the moderately compact
supermassive dark object that has been detected gravitationally,
and possibly also in X-rays
in the quiescent state,
with a radius of $\sim$ 20 mpc
\cite{bag},
could be two distinct
objects.

In summary,
using the Thomas-Fermi
theory, we have shown that
a  weakly interacting
self-gravitating
fermionic gas  at finite temperature
yields  a mass distribution that
successfully describes both the center and the halo
of the Galaxy.
For a fermion mass
$m \simeq 15$ keV,
a reasonable fit to the rotation
curve is achieved with the
temperature $T = 3.75$ meV and
the degeneracy parameter
at the center $\eta_0=28$.
With the same parameters,
the masses enclosed
within 50 and 200 kpc are
$M_{50} = 5.04\times 10^{11} M_{\odot}$
and
$M_{200} = 2.04\times 10^{12} M_{\odot}$,
respectively.
These values agree quite well with the mass estimates
based on the motion
of satellite galaxies and globular clusters
\cite{wilk}.
Moreover, the mass
 $M_{\rm c} \simeq 2.27\times 10^6 M_{\odot}$,
enclosed within 18 mpc,
agrees reasonably  well
with the observations of the compact dark object at
the center of the Galaxy.
We thus conclude that both the Galactic halo and the center
could be made of the same fermions.

%________________________
%{\bf Acknowledgement}
This
research is in part supported by the Foundation of Fundamental
Research (FFR) grant number PHY99-01241 and the Research Committee of
the University of Cape Town.
The work of N.B.\ is supported in part by
the Ministry of Science and Technology of the Republic of Croatia
under Contract No.\ 0098002.


\begin{thebibliography}{99}
%
\bibitem{1} M.A.\ Markov, Phys.\ Lett.\ {\bf 10}, 122 (1964).
%
\bibitem{2} R.D.\ Viollier, D.\ Trautmann and G.B.\ Tupper,
 Phys.\ Lett.\  {\bf B 306}, 79 (1993);
 R.D.\ Viollier,
 Prog.\ Part.\ Nucl.\ Phys. {\bf 32}, 51 (1994).
%
\bibitem{3} N.\ Bili\'{c}, D.\ Tsiklauri and R.D.\ Viollier,
 Prog.\ Part.\ Nucl.\ Phys.\ {\bf 40}, 17 (1998);
N.\ Bili\'{c} and R.D.\ Viollier,
 Nucl.\ Phys.\ (Proc.\ Suppl.) {\bf B 66}, 256 (1998).
%
\bibitem{4} N.\ Bili\'{c}, F.\ Munyaneza and R.D.\ Viollier,
Phys.\ Rev.\  {\bf D 59}, 024003 (1999).
%
\bibitem{5} D.\ Tsiklauri and R.D.\ Viollier,
Astropart.\ Phys.\ {\bf 12},
199 (1999);
 F.\ Munyaneza and R.D.\ Viollier,
 astro-ph/9907318.
%
\bibitem{6} F.\ Munyaneza, D.\ Tsiklauri and R.D.\ Viollier,
 Astrophys.\ J.\ {\bf 509}, L105 (1998);
{\it ibid}.\ {\bf 526}, 744 (1999);
F.\ Munyaneza and R.D.\ Viollier,
 Astrophys.\ J.\ {\bf 564}, 274 (2002).
%
\bibitem{kor}
L.C.\ Ho and J.\ Kormendy,
astro-ph/0003267;
astro-ph/0003268.
%
\bibitem{7} F.\ Macchetto {\it et al.},
 Astrophys.\ J.\ {\bf 489}, 579 (1997).
%
\bibitem{opp} J.R.\ Oppenheimer and G.M.\ Volkoff,
 Phys.\ Rev.\ {\bf 55}, 374 (1939).
%
\bibitem{eck} A.\ Eckart and R.\ Genzel,
 Mon.\ Not.\ R.\ Astron.\ Soc.\ {\bf 284}, 576 (1997);
 A.M.\ Ghez, B.L.\ Klein, M.\ Morris and E.E.\ Becklin,
 Astrophys.\ J.\ {\bf 509}, 678 (1998).
%
\bibitem{11} E.W.\ Kolb and M.S.\ Turner,
{\it The Early Universe}
(Addison-Wesley, San Francisco, 1989).
%
%
\bibitem{shi} X.\ Shi and G.M.\ Fuller,
 Phys.\ Rev.\ Lett.\ {\bf 82}, 2832 (1999);
K.\ Abazajian, G.M.\ Fuller, and M.\ Patel,
 Phys.\ Rev.\  {\bf D 64}, 023501 (2001);
G.B.\ Tupper, R.J.\ Lindebaum, and R.D.\ Viollier,
Mod.\ Phys.\ Lett.\  {\bf A 15}, 1221 (2000).
%
\bibitem{cov}
T.\ Goto and M.\ Yamaguchi,
 Phys.\ Lett.\  {\bf B 276}, 123 (1992).
%
\bibitem{lyt}
  D.H.\ Lyth,
 Phys.\ Lett.\  {\bf B 488}, 417 (2000),
%
\bibitem{col}
S.\ Cole and C.\ Lacey,
 Mon.\ Not.\ R.\ Astron.\ Soc.\ {\bf 281}, 716 (1996)
 and references therein.
%
%\bibitem{giu}
% G.F.\ Giudice, E.W.\ Kolb,
% A.\ Riotto,
%D.V.\ Semikoz and I.\ Tkachev,
% Phys.\ Rev.\  {\bf D 64}, 043512 (2001).
%
%\bibitem{13} M.\ Drees and D.\ Wright, hep-ph/0006274.
%
%\bibitem{14} P.\ de Bernardis {\it et al.},
% Nature {\bf 404}, 955 (2000).
%
%\bibitem{7} F.\ Macchetto {\it et al.},
%Astrophys.\ J.\ {\bf 489}, 579 (1997).
%
%\bibitem{15} S.\ Chandrasekhar, ``Stellar Structure''
%(University of Chicago Press, Chicago, 1939).
%
\bibitem{16} N.\ Bili\'{c} and R.D.\ Viollier,
 Phys.\ Lett.\  {\bf B 408}, 75 (1997).
%
\bibitem{bil1} N.\ Bili\'{c} and R.D.\ Viollier,
 Gen.\ Rel.\ Grav.\ {\bf 31},
1105 (1999).
%
\bibitem{bil2} N.\ Bili\'{c} and R.D.\ Viollier,
Eur.\ Phys.\ J.\  {\bf C 11}, 173 (1999).
%
\bibitem{17} W.\ Thirring, Z.\ Physik {\bf 235}, 339 (1970);
P.\ Hertel, H.\ Narnhofer and W.\ Thirring, Comm.\ Math.\ Phys.\
{\bf 28}, 159 (1972);
J.\ Messer,  J.\ Math.\ Phys.\ {\bf 22}, 2910 (1981).
%
\bibitem{lynd} D.\ Lynden-Bell,
 Mon.\ Not.\ R.\ Astron.\ Soc.\ {\bf 136}, 101 (1967).
%
\bibitem{shu} F.H.\ Shu, Astrophys.\ J.\ {\bf 225}, 83 (1978);
{\it ibid} {\bf 316}, 502 (1987).
%
\bibitem{kul} A.\ Kull, R.A.\ Treumann and H.\ B\"{o}hringer,
 Astrophys.\ J.\ {\bf 466}, L1 (1996).
%
\bibitem{padma} T.\ Padmanabhan,
{\it Structure formation in the Universe}
(Cambridge University Press, Cambridge, 1993).
%
\bibitem{bin} J.\ Binney and S.\ Tremaine, {\it Galactic Dynamics}
(Princeton University Press, Princeton, New Jersey, 1987).
%
\bibitem{wilk} M.I.\ Wilkinson and N.W.\ Evans,
 Mon.\ Not.\ R.\ Astron.\ Soc.\ {\bf 310}, 645 (1999).
%
\bibitem{bil3} N.\ Bili\'{c}, R.J.\ Lindebaum, G.B.\ Tupper, and
R.D.\ Viollier, Phys.\ Lett.\ {\bf B 515}, 105 (2001).
%
\bibitem{chav} P.-H.\ Chavanis and J.\ Sommeria,
 Mon.\ Not.\ R.\ Astron.\ Soc.\ {\bf 296}, 569 (1998).
%
\bibitem{you}
P.J.\ Young,
Astrophys.\ J.\ {\bf 81}, 807 (1976);
G.\ de Vaucouleurs and W.D.\ Pence,
Astrophys.\ J.\ {\bf 83}, 1163 (1978).
%
\bibitem{suc}
P.D.\ Sackett,
Astrophys.\ J.\ {\bf 483}, 103 (1997).
%
\bibitem{per} M.\ Persic,   P.\ Salucci, and F.\ Stell,
 Mon.\ Not.\ R.\ Astron.\ Soc.\ {\bf 281}, 27 (1986).
%
\bibitem{oll} R.P.\ Olling and M.R.\ Merrifield,
 Mon.\ Not.\ R.\ Astron.\ Soc.\ {\bf 311}, 361 (2000).
%
\bibitem{lo} K.H.\ Lo et al., Astrophys.\ J.\ {\bf 508}, L61 (1998).
%
\bibitem{zha} J.\ Zhao and W.M.\ Goss,
Astrophys.\ J.\ {\bf 499} L163 (1998).
%
\bibitem{bag} F.K.\ Baganoff {\it et al.}, Nature
 {\bf 413}, 45 (2001).
%\newpage
\end{thebibliography}
\end{document}